\documentclass[fleqn,usenatbib]{mnras}

\usepackage{newtxtext,newtxmath}

\usepackage[T1]{fontenc}
\usepackage{ae,aecompl}

\usepackage{graphicx}	
\usepackage{amsmath}	
\usepackage{amssymb}	
\usepackage{xcolor}
\usepackage{aas_macros}
\usepackage{hyperref} 
\hypersetup{colorlinks,citecolor=blue,linkcolor=blue,urlcolor=blue}

\hypersetup{draft}

\newcommand\rah{\mbox{$^{\mathrm h}$}}%
\newcommand\ram{\mbox{$^{\mathrm m}$}}%
\newcommand\arcdeg{\mbox{$^\circ$}}%

\title[UV Ceti with ASKAP]{ASKAP Detection of Periodic and Elliptically Polarized Radio Pulses from UV Ceti}

\author[Zic et al.]{Andrew Zic,$^{1,2}$\thanks{E-mail: azic7771@uni.sydney.edu.au}
Adam Stewart$^{1}$,
Emil Lenc$^{2}$,
Tara Murphy$^{1}$,
Christene Lynch$^{3,4}$,
\newauthor
David L. Kaplan$^{5}$,
Aidan Hotan$^{6}$,
Craig Anderson$^{6}$,
John D. Bunton$^{2}$,
\newauthor
Aaron Chippendale$^{2}$,
Stacy Mader$^{7}$,
Chris Phillips$^{2}$
\\
$^{1}$Sydney Institute for Astronomy, School of Physics, University of Sydney, NSW 2006, Australia.\\
$^{2}$CSIRO Astronomy and Space Science, PO Box 76, Epping, NSW 1710, Australia.\\
$^{3}$International Centre for Radio Astronomy Research - Curtin University, 1 Turner Avenue, Bentley, WA 6102, Australia.\\
$^{4}$ARC Centre of Excellence for All Sky Astrophysics in 3 Dimensions (ASTRO 3D).\\
$^{5}$Department of Physics, University of Wisconsin - Milwaukee, Milwaukee, Wisconsin 53201, USA.\\
$^{6}$CSIRO Astronomy and Space Science, 26 Dick Perry Avenue, Kensington, WA 6151, Australia.\\
$^{7}$CSIRO Astronomy and Space Science, Parkes Observatory, PO BOX 276, Parkes NSW 2870, Australia.\\
}

\date{Accepted XXX. Received YYY; in original form ZZZ}

\pubyear{2019}

\begin{document}
\label{firstpage}
\pagerange{\pageref{firstpage}--\pageref{lastpage}}
\maketitle

\begin{abstract}
    Active M-dwarfs are known to produce bursty radio emission, and multi-wavelength studies have shown that Solar-like magnetic activity occurs in these stars. However, coherent bursts from active M-dwarfs have often been difficult to interpret in the Solar activity paradigm. 
    We present Australian Square Array Pathfinder (ASKAP) observations of UV~Ceti at a central frequency of 888 MHz. We detect several periodic, coherent pulses occurring over a timescale consistent with the rotational period of UV~Ceti. The properties of the pulsed emission show that they originate from the electron cyclotron maser instability, in a cavity $\sim7$ orders of magnitude less dense than the mean coronal density at the estimated source altitude. These results confirm that auroral activity can occur in active M-dwarfs, suggesting that these stars mark the beginning of the transition from Solar-like to auroral magnetospheric behaviour. These results demonstrate the capabilities of ASKAP for detecting polarized, coherent bursts from active stars and other systems. 
\end{abstract}




\section{INTRODUCTION }
\label{sec:intro}
Magnetic reconnection events in stellar atmospheres drive powerful flare events causing sudden increases in luminosity from radio through to X-ray bands. Usually these occur in M dwarf stars, and the activity is caused by strong (kilogauss) magnetic fields \citep{JohnsKrull} that can produce flare energies up to 10$^3$ times that of the Sun \citep{Haisch}. Radio observations of flare stars were first performed in 1958 at megahertz frequencies \citep{Lovellfirst}, followed by a shift to gigahertz frequencies during the following decades. 

Below 5~GHz, radio emission from active M-dwarfs is dominated by coherent emission generated by instabilities in the electron velocity distribution \citep{Dulk, Bastian90, Melrose17}. This coherent emission can be used to probe features such as the structure of the electron velocity distributions \citep{Melrose93}, and other properties of the emission region such as the density and magnetic field strength \citep{Dulk,BenzandGudel}.

UV~Ceti is the prototype source of the flare star class (spectral type M5.5), and is part of a binary star system located at a distance of 2.7~pc \citep{gaia}. Radio emission from UV~Ceti was first reported by \cite{LovellUVCetiFirst} at a frequency of 240~MHz with flares lasting for 1--10s of minutes that occurred coincidentally with minor optical flares. Subsequent studies, mostly at megahertz frequencies, linked radio flares to flares seen at optical wavelengths, with radio emission following optical outbursts by tens of minutes \citep{LovellSimultaneous,LovellLargeFlare,Nelson}. 

With the advent of modern radio interferometers, observations at gigahertz frequencies became dominant in the following decades. Quiescent emission from UV~Ceti was detected at millijansky levels at 4.9~GHz, along with a fractional circular polarization value of 30--50~per~cent \citep{Linksy}. The first dynamic spectra of stellar flares other than those from the Sun were obtained by \cite{Bastian&Bookbinder} by observing two flares from UV~Ceti at 1.4~GHz, each lasting for approximately 10~minutes with peak flux densities of 200 and 100~mJy. The two coherent flares detected were highly circularly polarized (100 and 70~per~cent) with one flare showing evidence of variation within the 41~MHz bandwidth (also see \citealt{Jackson} for further dynamic spectra). 

The longest continuous interferometric observation of UV~Ceti was taken by \cite{Guedel96} for $\sim$9~h on two occasions at 4.9 and 8.5~GHz, with simultaneous coverage in soft X-rays. The authors found several weakly polarized radio events that began within a few minutes of X-ray flares and then peaked and decayed as the X-ray flares developed gradually, suggesting that the magnetic activity on M-dwarfs is analogous to Solar activity.

Most recently, \cite{Lynch} observed UV~Ceti at 154~MHz detecting four bursts each lasting $\sim$30~min and flux densities of between 10--65~mJy. These bursts were only detected in circular polarization (as well as linear polarization for the brightest burst), and have estimated brightness temperatures between 10$^{13}$--10$^{14}$\,K. These burst characteristics also led the authors to conclude that the emission was due to the electron cyclotron maser instability, and that the bursts may occur periodically on a time-scale consistent with the rotational period of UV~Ceti. \citet{Villadsen} have also recently reported the detection of several coherent bursts from UV~Ceti in wide-band dynamic spectra. Similarities in the complex morphologies of bursts detected months apart led \citet{Villadsen} to suggest that these bursts may be due to periodic pulses of electron cyclotron maser emission. Coherent bursts from active M-dwarfs have proven difficult to interpret within the paradigm of Solar activity \citep{Bastian90, Villadsen}, but these recent results suggest that auroral processes, prevalent in brown dwarfs \citep{Pineda} and magnetized planets \citep{Zarka98} may also be relevant to these stars.

In this paper we present results of two $\sim$10~hr observations of UV~Ceti, separated by $\sim$5~months, at 888\,MHz using the Australian Square Kilometre Array Pathfinder telescope \citep[ASKAP;][]{Johnston08, askap}. We demonstrate the variability and polarization capabilities of the telescope by analysing the dynamic spectra of each observation over the 288\,MHz ASKAP bandwidth, where we detect periodic coherent bursts associated with UV~Ceti recurring on timescales consistent with the rotational period of the star. We investigate the bursts to determine the emission mechanism, and conclude that  rotationally-modulated (i.e. pulsed) emission from the electron cyclotron maser instability is the most likely origin. We argue that the time-frequency structure and polarization properties of the pulses show that they originate along localised magnetic loops within extreme density cavities in the magnetosphere of UV~Ceti.

\cite{Villadsen} showed that the duty cycle of luminous coherent bursts from active M-dwarfs peaks at 25\,per~cent at 1--1.4\,GHz, a frequency range that is fully observable with ASKAP. This highlights the potential for numerous detections of coherent bursts from active M-dwarfs in future surveys such as the ASKAP Variables and Slow Transients (VAST) survey \citep{Murphy13}.The excellent data quality from ASKAP will enable detailed analyses of these bursts, allowing these surveys to probe the magnetospheric processes occurring in the population of active M-dwarfs.


\section{OBSERVATIONS \& DATA REDUCTION}
\label{sec:datareduction}
We observed UV~Ceti on two epochs (2018-10-02 and 2019-03-07) during ASKAP commissioning tests, with details summarised in Table~\ref{obsdetails}. 
\begin{table*}
\caption{The details of the observations of UV~Ceti with ASKAP used in this work. `C. Freq.' is the central frequency of the observation and `BW' is the bandwidth.} 
\centering
\begin{tabular}{cccccc}
\hline
Obs. Date & Duration & C. Freq. & BW & Integration & Num. of \\
(UTC) & (h) & (MHz) & (MHz) & Time (s) & Antennas \\
\hline
2018 Oct 02 12:02 & 10 & 888 & 288 & 10 & 28 \\
2019 Mar 07 01:41 & 10.5 & 888 & 288 & 10 & 33 \\
\hline
\end{tabular}
\label{obsdetails}
\end{table*}

\begin{figure*}
	\includegraphics[width=\linewidth]{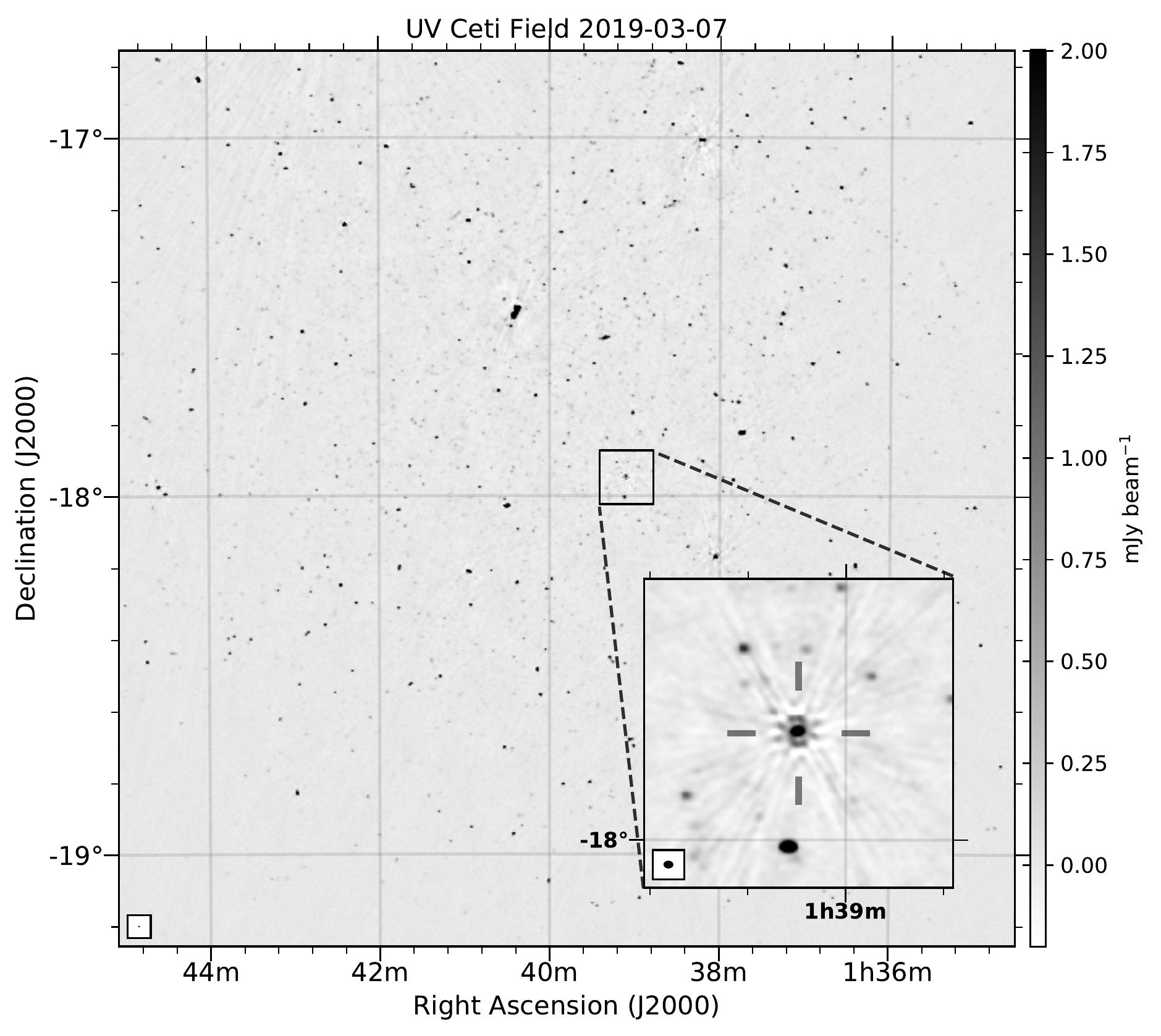}
    \caption{Image showing the continuum field containing UV~Ceti as observed by ASKAP on 2019-03-07. Note that UV~Ceti, which is shown centered in the `zoom-in` section of the map, has been excluded from the deconvolution process. The synthesised beam is $17.4\arcsec\times13.6\arcsec$ FWHM at a position angle of $88\arcdeg$. The image peak flux density is 342 mJy beam$^{-1}$ and the image RMS noise is $24$\,$\mu$Jy\,beam$^{-1}$.}
    \label{fig:field}
\end{figure*}


We reduced the data using the Common Astronomy Software Applications package \citep[\textsc{casa},][]{casa}. We observed the primary calibrator PKS B1934$-$638 prior to each observation to calibrate the flux scale, the instrumental bandpass, and polarization leakage. On-dish calibrators were used to calibrate the frequency-dependent $XY$-phase for the 2019-03-07 observation, allowing us to correct for leakage between Stokes $Q$ and $V$. The observation on 2018-10-02 was not configured to allow correction of frequency-dependent $XY$-phase, which results in substantial leakage between Stokes $Q$ and Stokes $V$ for this observation.
We performed basic flagging to remove radio-frequency interference that affected approximately 6~per~cent of the visibility data. We shifted the phase centre of the 2018-10-02 visibility data to the pointing centre for beam 20 (RA$=$01\rah39\ram47\fs742 and Dec$=$-18\arcdeg00\arcmin30\farcs34) using the task \textsc{fixvis}. This was not needed for the 2019-03-07 observation, because we enabled fringe tracking per beam. For each observation, we performed an initial shallow stage of deconvolution using the \textsc{clean} task (with briggs weighting and a robustness of 0.5) to assess phase stability. 
We used the task \textsc{gaincal} to perform phase-only self-calibration. The resulting phase corrections were stable as a function of time with only slow variations associated with the longest baselines.

We constructed a mask excluding a $1.5\arcmin$ square region centred on UV~Ceti using the initial field image. This allowed subsequent deconvolution to model the field sources without removing the time- and frequency-dependent effects of UV~Ceti. We then performed a deep clean (20000 iterations) using this mask. We used the multi-scale algorithm (with scales of 0, 5, and 15 pixels and a cell size of 2.5\arcsec) to account for slightly extended sources and two Taylor terms to model sources with non-flat spectra. We excluded baselines shorter than $200\lambda$ ($\sim70$\,m) to avoid contamination from diffuse emission.
Figure \ref{fig:field} shows the restored image of the 2019-03-07 observation that achieves a sensitivity of 24\,$\mu$Jy\,beam$^{-1}$ with a $17.4\arcsec\times13.6\arcsec$ restoring beam. UV~Ceti is visible in the continuum image with a peak flux density of 6.3\,mJy\,beam$^{-1}$.

To generate dynamic spectra for UV~Ceti, we subtracted the Stokes I field model from the Stokes I visibilities with the \textsc{casa} task \textsc{uvsub} and then phase rotated from the beam centre to the location of UV~Ceti at the respective epochs based on GAIA DR2 \citep[RA$=01\rah39\ram05\fs8156$ and Dec$=-17\arcdeg56\arcmin50\farcs021$,][]{gaia} with the task \textsc{fixvis}. We baseline-averaged the subtracted visibilities, using only baselines greater than $200$\,m to avoid contamination from diffuse emission, for each of the instrumental polarizations. 

To produce dynamic spectra for the four Stokes parameters $(I, Q, U, V)$, we combined the complex visibilities for each instrumental polarization as follows:
\begin{align}
    I &= {(XX + YY)}/{2}\\
    Q &= -{(XY + YX)}/{2}\\
    U &= -{(XX - YY)}/{2}\\
    V &= j{(XY - YX)}/{2},
\end{align}
where $(XX, YY, XY, YX)$ are the instrumental polarizations, and $j$ is the imaginary unit, $\sqrt{-1}$. 

The unconventional definitions of Stokes parameters given above arise due to the 45 degree angle of the $X$ and $Y$ feeds relative to the vertical axis at transit \citep{Sault_pol} and other instrumental factors revealed during commissioning. With the above definitions, the Stokes parameters follow the IAU standard of polarization. We verified the polarization properties of the array with a short observation of the Vela pulsar with the same correlator settings as used in the UV~Ceti observations.

Due to the lack of $XY$-phase calibration for the 2018-10-02 epoch, we can only consider Stokes $Q$ and $V$ in quadrature for this observation, i.e. $\sqrt{Q^2 + V^2}$. Full polarimetric calibration (leakage and frequency-dependent $XY$-phase) for the observation on 2019-03-07 was possible, which enabled us to generate dynamic spectra for each Stokes parameter. We also generated dynamic spectra of the total fractional (elliptical) polarization by computing
\begin{equation}
    f_p = \frac{\sqrt{Q^2 + U^2 + V^2}}{I}
\end{equation}
for the dynamic spectra from both observing epochs. To avoid spuriously high polarization fractions generated by noise in the dynamic spectra, we masked any values corresponding with Stokes I flux densities below a $4\sigma$ threshold. 

To increase the signal to noise in our dynamic spectra, we averaged over 4 frequency channels, giving our dynamic spectra a resolution of 10\,s in time, and 4\,MHz in frequency. We calculated the RMS noise in our dynamic spectra by taking the standard deviation of their imaginary parts for each Stokes parameter. The average RMS noise of the dynamic spectra across all Stokes parameters was 9.7\,mJy for 2018-10-02, and 7.4\,mJy for 2019-03-07.

\section{RESULTS}

\begin{figure}
\begin{center}
	\includegraphics[width=\linewidth]{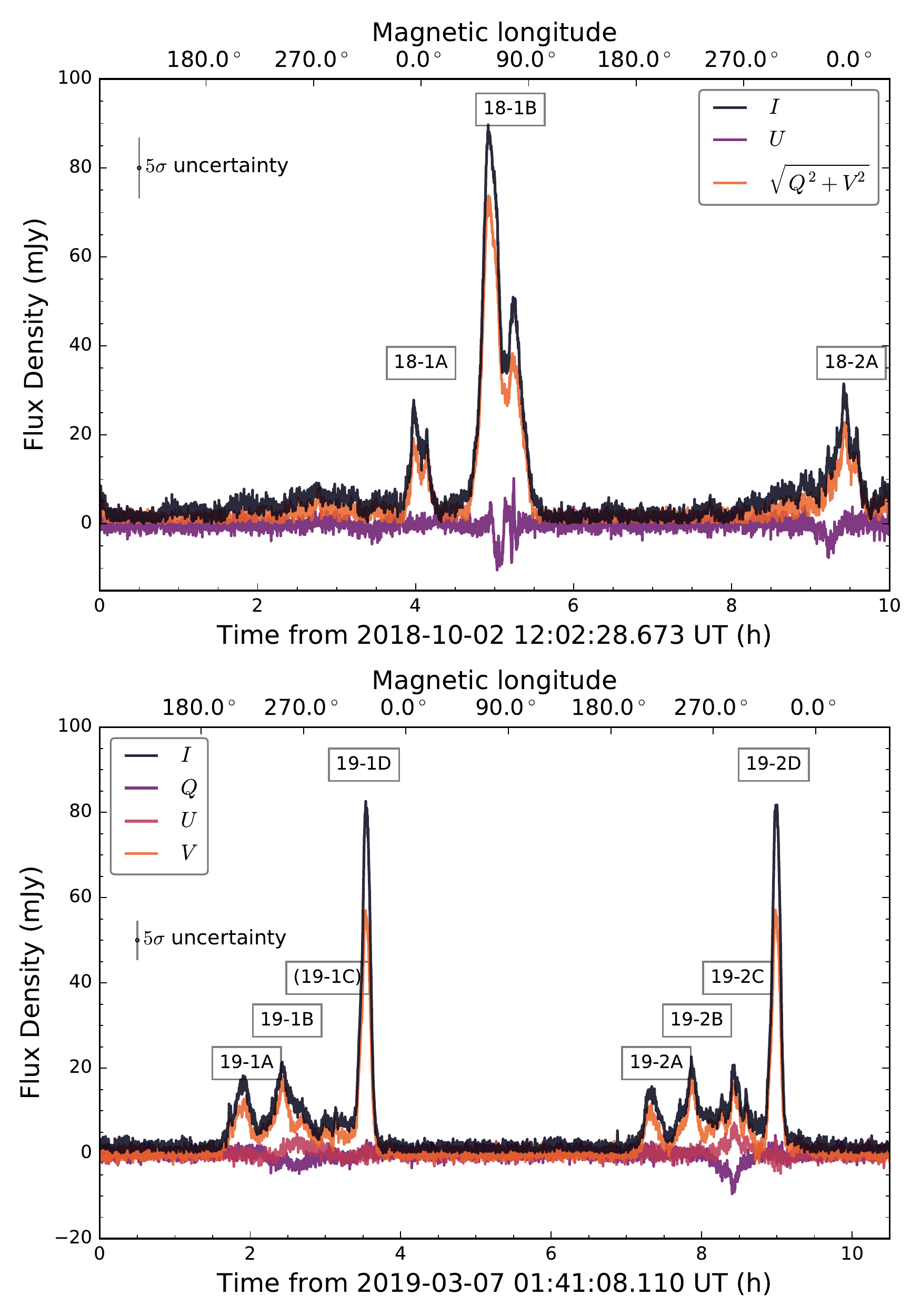}
    \caption{Frequency-averaged light curve for UV~Ceti in the 2018-10-02 epoch (top panel) and 2019-03-07 epoch (bottom panel). Sub-structure in total intensity and in the Stokes parameters are evident in the pulses. Typical $5\sigma$ uncertainties are labelled in each plot. The lower abscissae show the time in hours from the observation start, and the upper abscissae show the relative magnetic longitude, as described in Section \ref{sec:compare_bursts}.}
    \label{fig:uvceti_lc}
    \end{center}
\end{figure}


The dynamic spectra for the 2018-10-02 and 2019-03-07 epochs are shown in Figures \ref{fig:ds_2018} and \ref{fig:ds_2019} respectively, and their frequency-averaged light curves are shown in Figure \ref{fig:uvceti_lc}. To measure the flux density uncertainties for each 10\,s light curve sample, we took the standard deviation of the imaginary part of the complex light curve. We detect several highly polarized pulses in both observing epochs, with a peak flux density in the dynamic spectra of $160\pm10$\,mJy in the 2018-10-02 epoch, and $139.7\pm7.6$\,mJy in the 2019-03-07 epoch.



\subsection{Emission Properties}

\subsubsection{\label{sec:burststructure1}2018-10-02: Temporal and spectral structure}

\begin{figure*}
    \begin{center}
	\includegraphics[width=\linewidth]{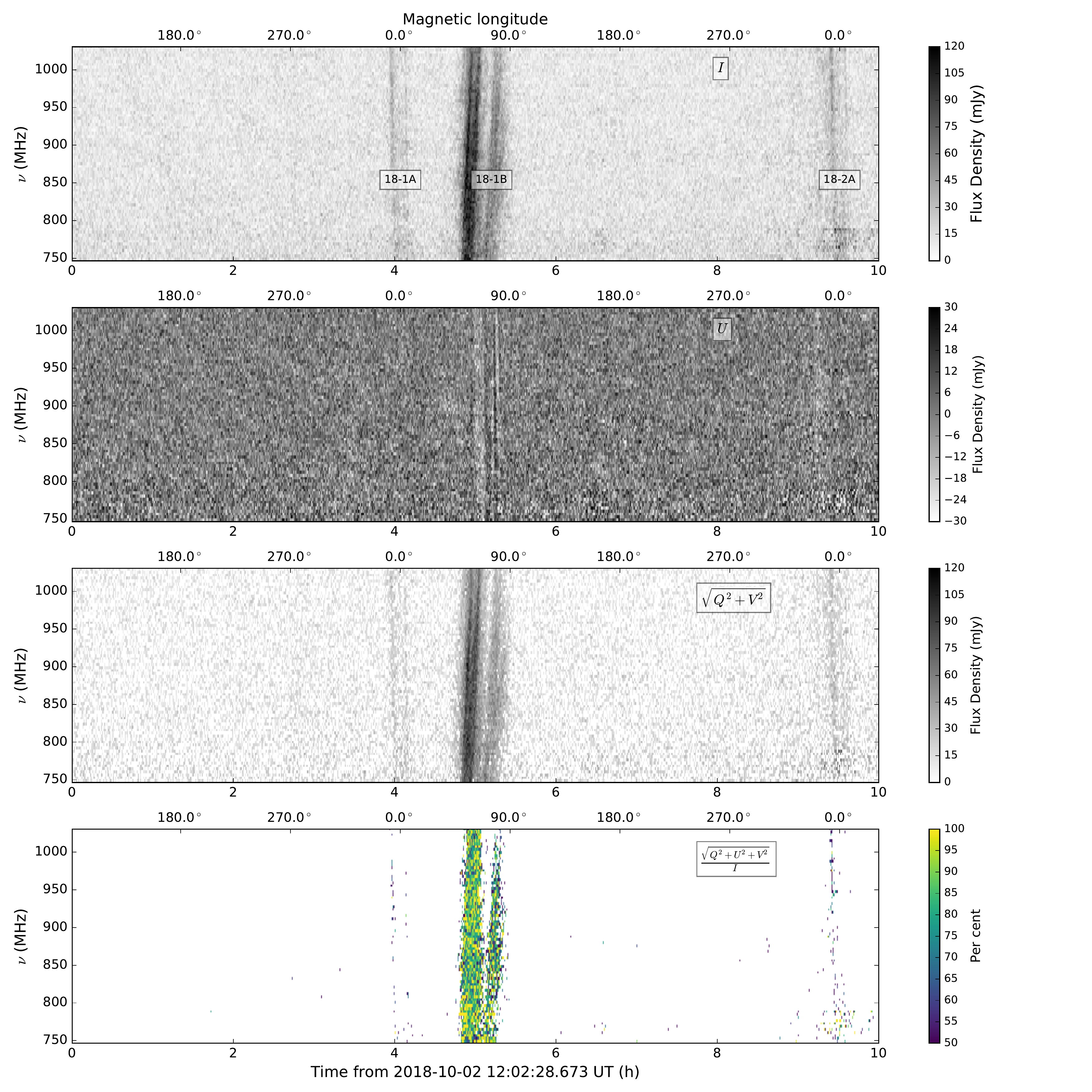}
    \caption{Dynamic spectrum of UV~Ceti in Stokes $I$, $U$, $\sqrt{Q^2 + V^2}$, and total fractional polarization $\sqrt{Q^2 + U^2 + V^2}/I$. The dynamic spectra have been averaged to 4\,MHz in frequency to improve the signal-to-noise. Total fractional polarizations that correspond with Stokes I flux densities below a $4\sigma$ have been blanked. To aid clarity, the pulses have been labelled 18-1A, 18-1B, and 18-2A. The similar drift rate and flux density of pulses 18-1A and 18-2A is evident. As in Figure \ref{fig:uvceti_lc}, the lower abscissae show the time in hours from the observation start, and the upper abscissae show the relative magnetic longitude, as described in Section \ref{sec:compare_bursts}.}
    \label{fig:ds_2018}
    \end{center}
\end{figure*}

\begin{figure}
    \begin{center}
	\includegraphics[width=\linewidth]{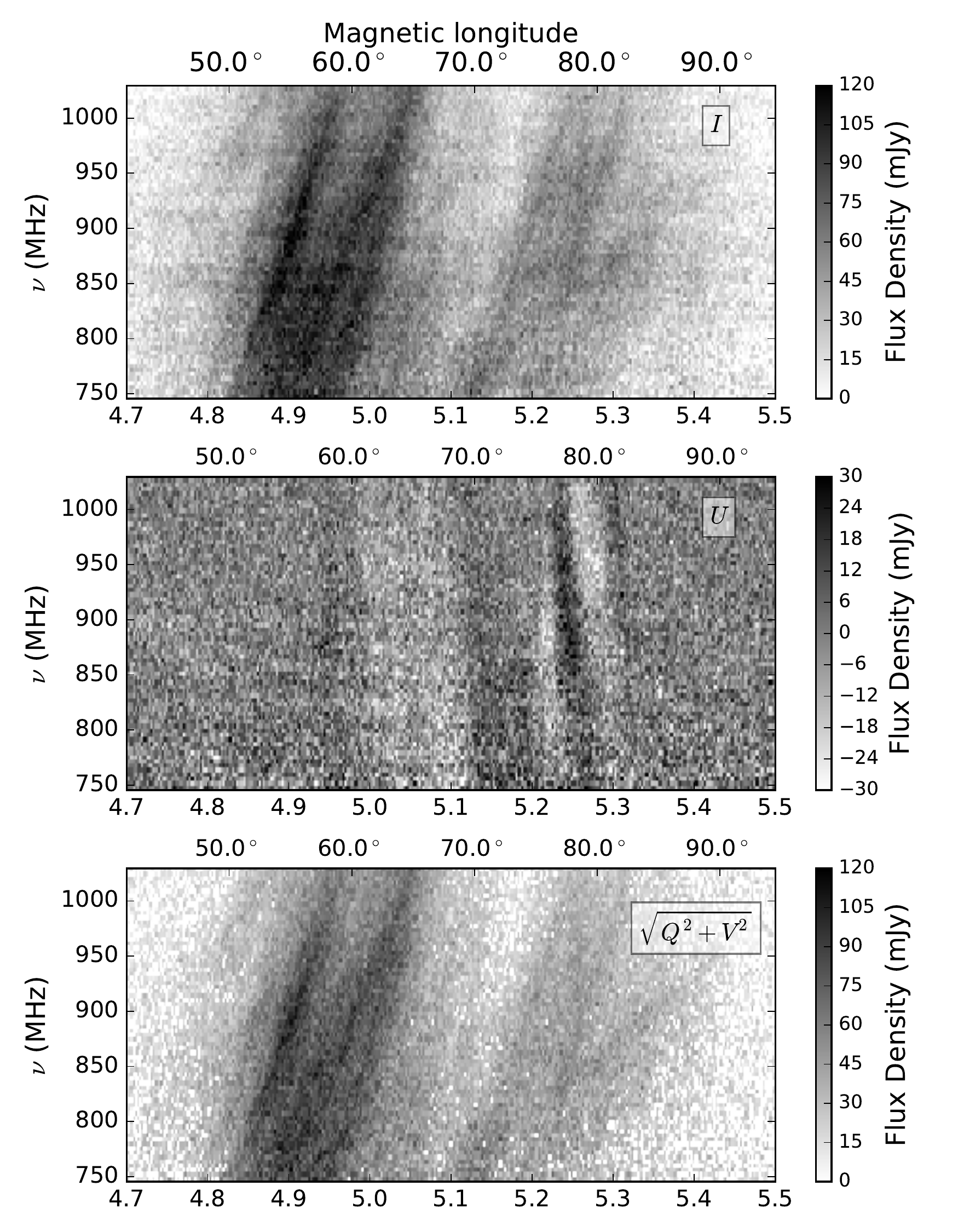}
    \caption{Detail of the pulse 18-1B in the 2018-10-02 epoch, in Stokes $I$, $U$, and $\sqrt{Q^2 + V^2}$. Complex sub-structure in total intensity, and in polarization is evident. Particularly notable is the banding present in Stokes $Q$, which appears to be drifting in an opposite direction to the bulk frequency drift of the pulse. As in Figure \ref{fig:ds_2018}, upper abscissae show the relative magnetic longitude.}
    \label{fig:2018_DS_IQ}
    \end{center}
\end{figure}

We detected three pulses in the 2018-10-02 epoch: a faint pulse with a peak flux density of $59\pm10$\,mJy (labelled 18-1A), followed by a $160\pm10$\,mJy pulse occurring $\sim 45$\,min after the first faint pulse (labelled 18-1B). Another $70\pm10$\,mJy pulse (labelled 18-1C) occurs $\sim5.4$\,h after 18-1A. In addition, the light-curve appears to show the tail end of a pulse that peaked just before the observation started, and about 5.4 hours before the tail of 18-1B. Hereafter, we refer to the pulses by their labels, which are shown in the Stokes I panel of Figure \ref{fig:ds_2018}. 

All three fully-detected pulses have a double-peaked profile, with more complex sub-structure evident within the dynamic spectrum of 18-1B. There is a gradual increase in flux density from $t\sim8$\,h until 18-2A peaks at $t\sim9.4$\,h. Preceding 18-1A, there is only a marginal flux density enhancement at $t\sim2.8$\,h. Apart from the preceding behaviour, 18-1A and 18-2A show a similar temporal profile. 

All three pulses exhibit a bulk frequency drift across the bandwidth of our observations. To determine the bulk frequency drift rates of the pulses, we fit a simple model of a series of drifting top-hat pulses to the observed dynamic spectra using a Markov Chain Monte Carlo (MCMC) sampler (\textsc{emcee}\footnote{\url{http://dfm.io/emcee/}}; \citealp{dfm}). We set uniform priors on the parameters of each pulse (central time, duration, amplitude, drift rate, and periodicity), within conservative large ranges determined by visual inspection.


We found that 18-1B was best fit by two components: the earlier component with a drift rate of $ \dot{\nu} = 0.75\pm0.03\,\mathrm{MHz}\,\mathrm{s}^{-1}$, and the later component with $\dot{\nu} = 0.60^{+0.03}_{-0.02}\,\mathrm{MHz}\,\mathrm{s}^{-1}$. We find that 18-1A and 18-2A have consistent drift rates of $-1.30^{+0.03}_{-0.02}\,\mathrm{MHz}\,\mathrm{s}^{-1}$ and $-1.30 \pm 0.03\,\mathrm{MHz}\,\mathrm{s}^{-1}$ respectively.

\subsubsection{\label{sec:burststructure2}2019-03-07: Temporal and spectral structure}

\begin{figure*}
    \begin{center}
	\includegraphics[width=\textwidth]{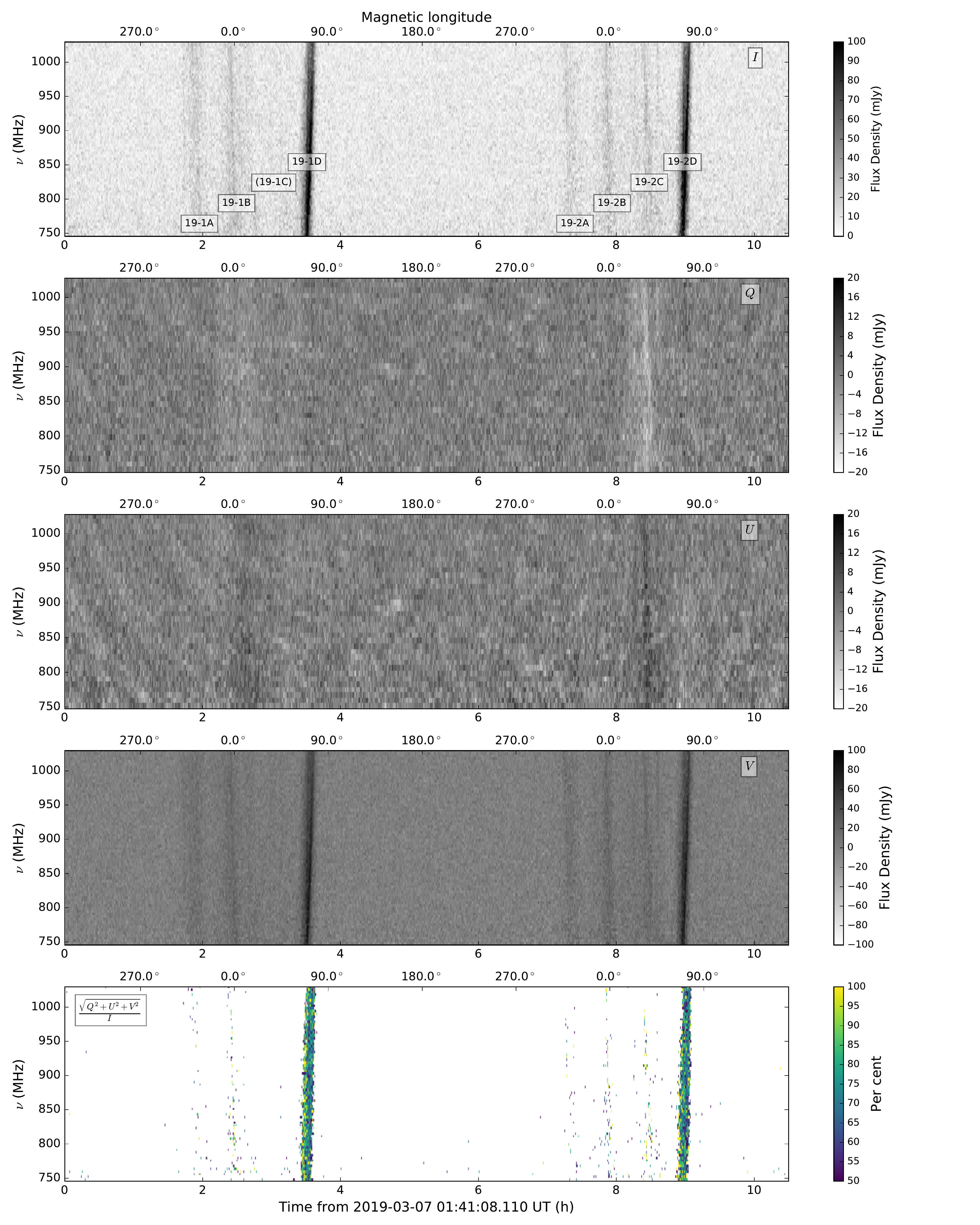}
    \caption{Full-Stokes dynamic spectrum of UV~Ceti. The dynamic spectra have been averaged to 4\,MHz in frequency in Stokes $I$ and $V$, and to 60\,s and 16\,MHz in Stokes $Q$ and $U$ to improve sensitivity. The bottom panel shows total fractional polarization similarly to Figure \ref{fig:ds_2018}. The pulses have been labelled 19-1A, 19-1B, (19-1C), 19-1D, 19-2A, 19-2B, 19-2C, 19-2D. Pulse (19-1C) is within brackets to indicate its marginal status, but becomes evident after further averaging of the dynamic spectrum. Upper abscissae show relative magnetic longitude as described in Section \ref{sec:compare_bursts}.}
    \label{fig:ds_2019}
    \end{center}
\end{figure*}

\begin{figure*}
    \begin{center}
	\includegraphics[width=\linewidth]{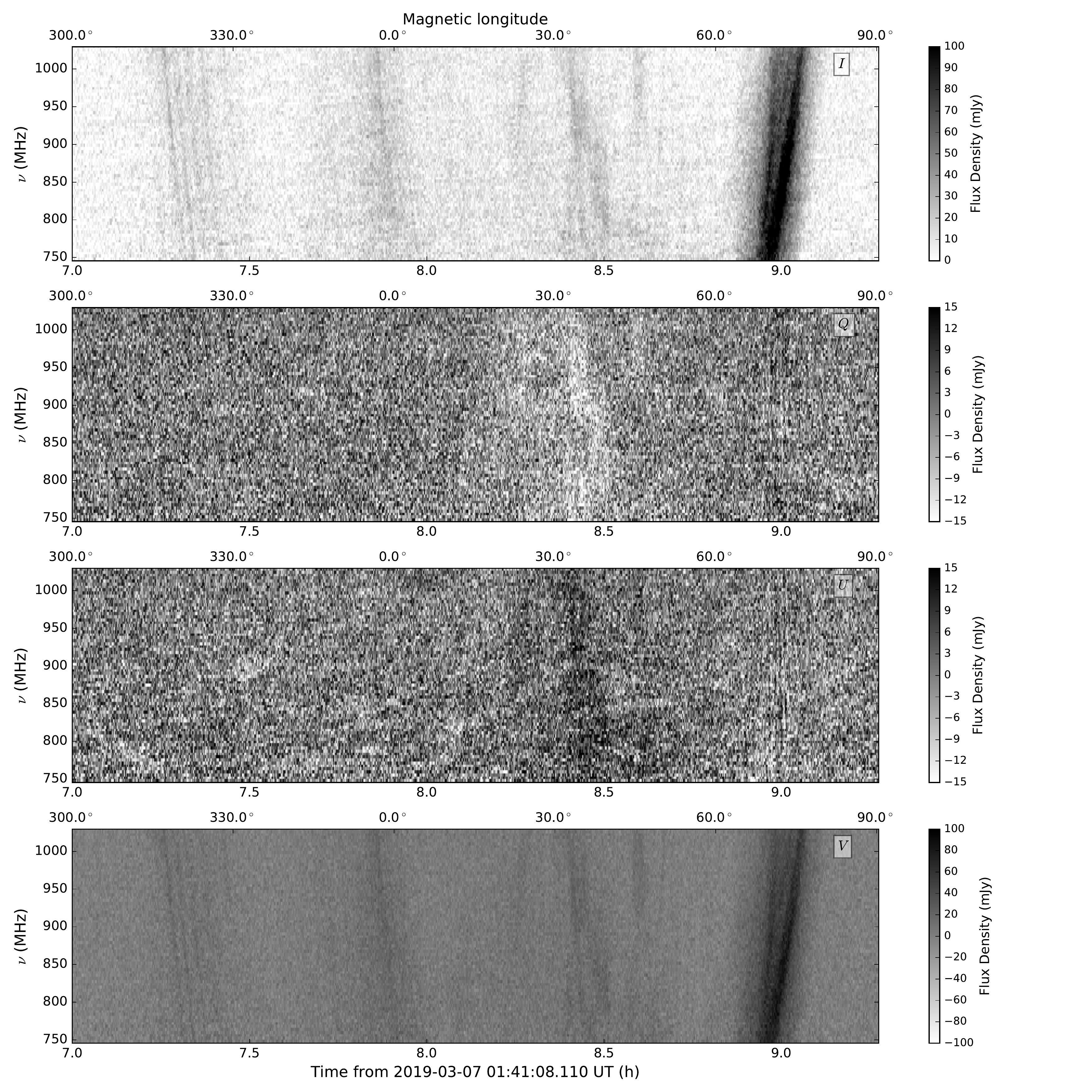}
    \caption{Detail of the pulses 19-2A, 19-2B, 19-2C, and 19-2D in Stokes $I$, $Q$, $U$, and $V$. The varying sub-structure, but similar overall drift rate of 19-2A, B, and C is evident. Pulse 19-2C exhibits linear linear polarisation with little dependence on frequency. As in Figure \ref{fig:ds_2019}, upper abscissae show relative magnetic longitude.}
    \label{fig:ds_2019_substructure}
    \end{center}
\end{figure*}


We detect eight clearly separated pulses in the 2019-03-07 epoch. These pulses appear to be within two distinct sets separated by about $5.4$\,h. We label the pulses in the first set as 19-1A, 19-1B, (19-1C), 19-1D, and 19-2A, 19-2B, 19-2C, 19-2D in the second set. Each set of pulses has three faint pulses with peak flux densities ranging from $16.2-51.6\pm7.6$\,mJy, followed by a bright pulse with a peak flux density of $136.4\pm7.6$\,mJy in the first set, and $139.7\pm7.6$\,mJy in the second set. (19-1C) is not clearly detected in the high-resolution dynamic spectra, but becomes apparent after further averaging in time and frequency. We label it within brackets to indicate its marginal status.

Each of the three faint pulses in each set shows different temporal structure to each other. There is some change in the temporal structure of 19-1A, C and their corresponding pulses 19-2A, C, although 19-1B appears to maintain a more steady, sharply-peaked temporal profile in the corresponding pulse 19-2B. The faint pulses in each set (19-1A, B, C and 19-2A, B, C) appear to be regularly separated, by approximately $32$\,min. 

Following the MCMC procedure outlined in Section \ref{sec:burststructure1}, we measured the drift rates and other parameters of the pulses. All faint pulses (19-1A, B, C and 19-2A, B, C) show a drift rate consistent with $-1.300\,\mathrm{MHz}\,\mathrm{s}^{-1}$, with typical uncertainties in the range of $0.003-0.005$\,MHz\,s$^{-1}$. Pulse 19-2A shows quasi-periodic striations that repeat approximately once every $1.8-1.9$\,min, similar to quasi-periodic oscillations from active M-dwarfs reported by \citet{Gary82, Lang86}, and \citet{BastianDS}. The oscillations in 19-2A are evident in Figure \ref{fig:ds_2019_substructure}.

Pulses 19-1D and 19-2D are very similar in their peak flux densities, and in their temporal sub-structure, both containing two closely-spaced components with slightly differing frequency drift rates. The bulk drift rates of 19-1D and 19-2D are both consistent with a value of $1.000\,\mathrm{MHz}\,\mathrm{s}^{-1}$, with uncertainties in the range of $0.002-0.004$\,MHz\,s$^{-1}$.

\subsubsection{\label{sec:polarization}Polarization}
In each epoch, the detected pulses are highly circularly polarized, averaging about 70 per cent circular polarization and reaching up to 100 per cent. Unfortunately due to the lack of cross-hand phase calibration in the 2018-10-02 epoch, determining the sense of circular polarization is difficult. However, in the 2019-03-07 epoch, the Stokes $V$ emission is consistently negative-valued, indicating that it is RCP. This is consistent with the majority of coherent bursts reported from UV~Ceti in the literature (e.g. \citealp{Villadsen, Lynch, Bastian&Bookbinder}); however, some LCP bursts have been reported \citep{Lynch}.

We detect linearly polarized emission associated with some of the pulses, at an average level of about 15 per cent of the total intensity. Inspection of the Stokes $U$ dynamic spectrum shows two linearly-polarised components in pulse 18-1B. The first Stokes $U$ component of 18-1B is largely positive-valued, exhibiting very little dependence on frequency. The second component in Stokes $U$ exhibits striations drifting in frequency in an opposite direction to the enveloping pulse. This behaviour is most evident in Figure \ref{fig:2018_DS_IQ}. There is also a weak, negative Stokes $U$ signal associated with 18-2A near $t = 9.4$\,h that also shows little dependence on frequency. 

In the 2019-03-07 epoch, we detect weak linearly polarized emission associated with pulses 19-1B and 19-2C. Like the linear polarization in 18-2A, the linear polarization in 19-1B and 19-2C show little dependence on frequency. There is also some marginal, rapidly-striating linear polarization associated with pulses 19-1D and 19-2D (evident in Figures \ref{fig:ds_2019} and \ref{fig:ds_2019_substructure}).

The detection of pulses in Stokes $Q$ and $U$ along with $V$ implies that the emission is elliptically polarized -- a rare occurrence in nature that is generally only seen in pulsars (e.g. \citealp{Melrose17}) and magnetized planets (e.g. \citealp{Zarka98, JupiterDAM, Boudjada, fischer}). We are aware of two other reports of elliptically polarized stellar radio emission from the literature: \citet{Lynch} also detected elliptically polarized emission from UV~Ceti with the Murchison Widefield Array at 154\,MHz, and \citet{Spangler74} detected elliptically polarized emission from AD~Leo at 430\,MHz.

The differing structure of the linearly polarized emission compared with the Stokes $I$ and $V$ emission strongly suggests that these polarization components are not instrumental in origin. Nonetheless we conducted a number of checks to verify that the linear polarization is legitimate.

We inspected bright sources around the UV Ceti field, and found that in general, within 40\arcmin{} of the field centre (corresponding with the half-width at half-maximum of the primary beam), leakage from Stokes $I$ to $V$ is $\sim 0.1$ per~cent, $Q$ to $I$ is $\lesssim 0.5$\,per~cent, and $U$ to $I$ is $\lesssim 1.5$\,per~cent. Towards the pointing centre, the leakage values decrease below the noise floor. These leakage fractions are at least 10 times lower than the measured degree of linear polarization from UV~Ceti ($\sim 15$\,per~cent). We obtain similar results with the short Vela observation reported above.

We also imaged the UV~Ceti field during times of apparent linearly polarized emission. These images show a linearly-polarized point source associated with UV~Ceti, and show that the emission is not an artefact associated with radio-frequency interference, or sidelobe confusion from a bright field source. At these times, other field sources show no spurious deviations from their baseline leakage values reported above. With the results of these tests, we are confident that the linearly polarized emission is intrinsic to UV~Ceti, and not instrumental in origin.

\begin{figure*}
    \begin{center}
	    \includegraphics[width=\linewidth]{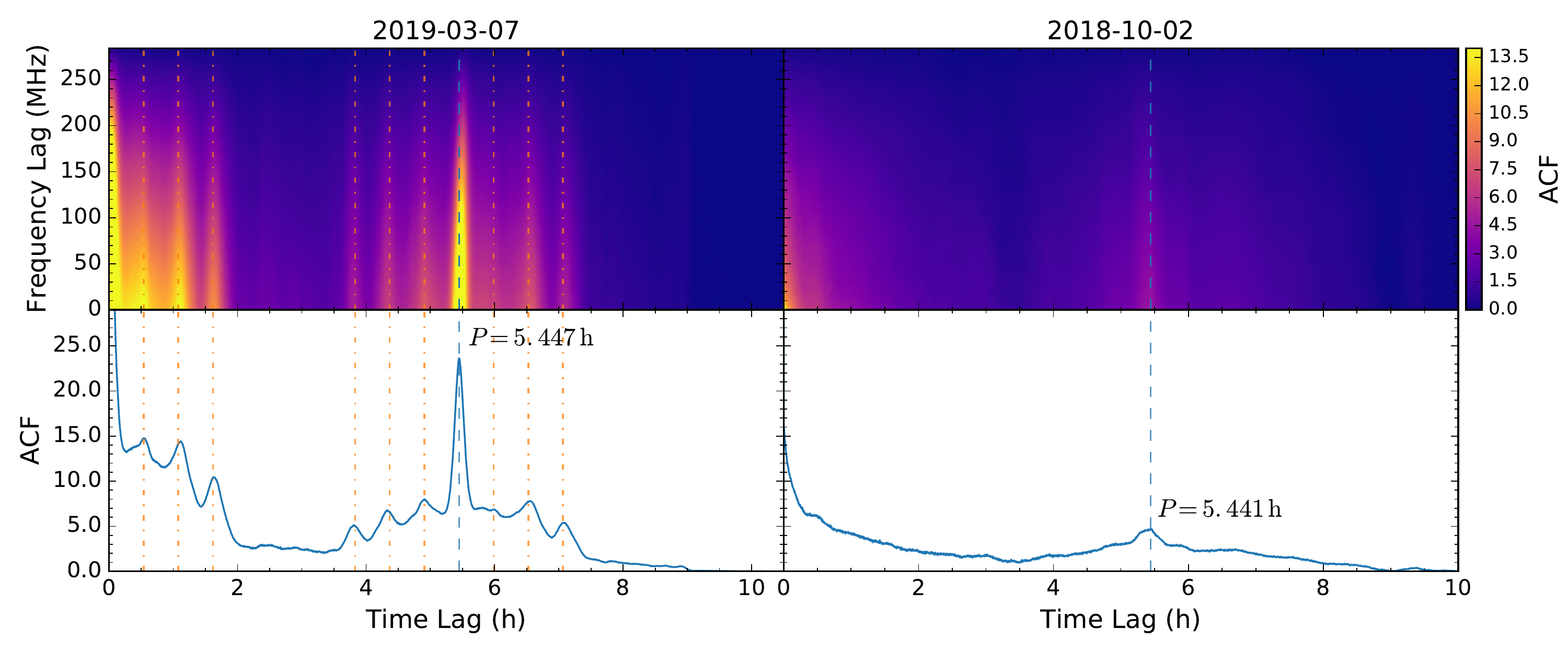}
    \caption{Two-dimensional autocorrelation function, and the corresponding zero frequency-lag components for the 2019-03-07 (left panels) and 2018-10-02 epochs (right panels). Dashed lines indicate the time lag of the major peak in the autocorrelation, which measures the periodicity of the pulses. The dash-dotted lines in the autocorrelation of 2019-03-07 are spaced every $0.539\,$h from the zero time lag component, and around the major peak at a time lag of $5.447$\,h, and show the quasi-periodic nature of the pulses 19-1A, B, C and 19-2A, B, and C.}
    \label{fig:acf}
    \end{center}
\end{figure*}

\subsection{\label{sec:burst_period}Derived Pulse Period}
The repetitive nature of the detected emission over the rotational period of UV Ceti strongly suggests that the emission is rotationally-modulated, pulsed, emission from highly-beamed radio sources, as opposed to stochastic bursts or flares.
To test the periodicity of the pulses, we computed the two-dimensional autocorrelation function of the Stokes I dynamic spectra, and measured the time lag at the peak along the zero frequency-lag component of the autocorrelation. 

The 2018-10-02 observation is dominated by the central bright burst, which does not repeat in our observation. To avoid this dominating the autocorrelation function, we masked the region from $t = 4.6 - 5.8$\,h and replaced it with artificial complex Gaussian noise with the same properties as measured in the region of the dynamic spectrum with no detectable emission $t = 0.3-1.5$\,h.
The resulting autocorrelation function (right panels of Figure~\ref{fig:acf}) shows a significant peak at $5.441\pm0.035$\,h, in the zero frequency-lag component. Similarly, the autocorrelation of the (unfiltered) 2019-03-07 epoch (left panels of Figure \ref{fig:acf}) shows a strong zero frequency-lag peak centred at $5.447\pm0.008$\,h. Note that the quoted uncertainties are estimates based on the sample distribution of periods from the simple top-hat pulse model described in Sections \ref{sec:burststructure1} and \ref{sec:burststructure2}. 

The pulse periods measured between the two epochs are consistent with one another, and consistent with the $5.4432\pm0.0072$\,h period measured by \citet{UVCetRotation}. We note that this disambiguates which component of the L726-8AB binary the emission originates from -- the primary component BL~Ceti (L726$-$8A) is also known to be active at radio frequencies (e.g. \citealp{Gary82}), and is separated from UV~Ceti by $\sim 4\arcsec{}$, well below the $\sim 15\arcsec{}$ resolution of our ASKAP observations. However, BL~Ceti has a rotational period of $5.832\pm 0.012$\,h \citep{UVCetRotation}.

Along with the $5.447\pm0.008\,$h periodicity revealed by the autocorrelation of the 2019-03-07 observation, there are also three minor peaks in the autocorrelation spaced every $0.539\,$h from the zero time-lag component, and to either side of the primary period of $5.447\pm0.008\,$h. This gives strong evidence for the quasi-periodic nature of the faint pulses 19-1A, B, C and 19-2A, B, C.

\subsection{\label{sec:compare_bursts}Comparison of the two epochs}

\begin{figure*}
    \begin{center}
	\includegraphics[width=\linewidth]{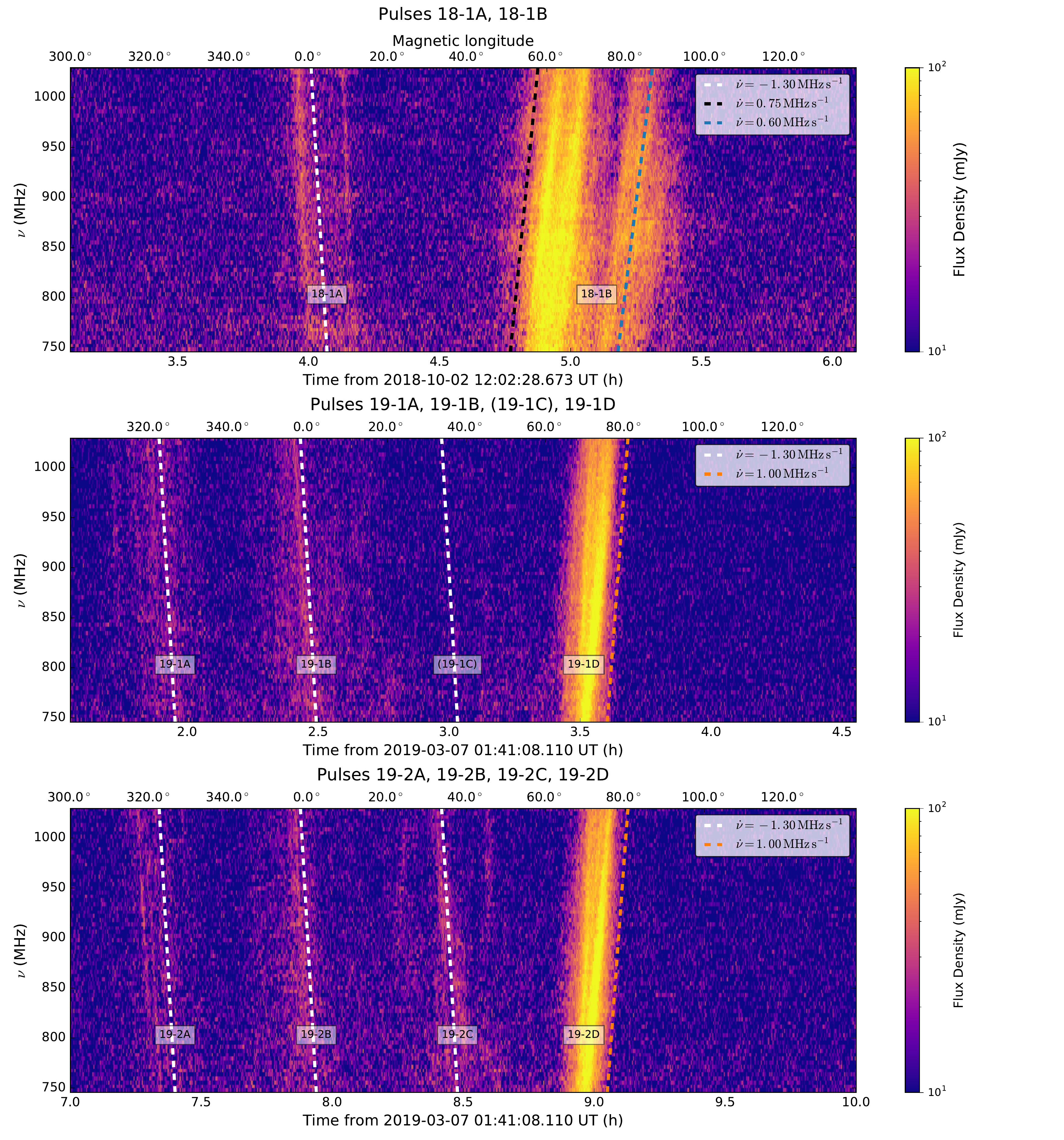}
    \caption{\label{fig:compare_bursts}Comparison of pulses from 2018-10-02 and 2019-03-07, at the same magnetic longitude (rotational phase). The pulses are labelled according their labels designated in Sections \ref{sec:burststructure1} and \ref{sec:burststructure2}. The dashed lines indicate the locations and drift rates of the pulses. The white dashed lines are spaced every $0.539\,$h in the 2019-03-07 panels, consistent with the periodicity of these pulses demonstrated by the autocorrelation function shown in Figure \ref{fig:acf}. The similar structure and arrangement of the pulses between the 2018-10-02 and 2019-03-07 epochs is evident. Upper abscissae show relative magnetic longitude for each of the pulses, assuming a fiducial zero-point at the peak of 19-1/19-2B, and 18-1A. To highlight pulse structure on different flux density scales, the colormap has been logarithmically scaled between $10-100$\,mJy.}
    \label{fig:compare_bursts}
    \end{center}
\end{figure*}

We find notable similarities between the pulses occurring in both epochs when inspecting the dynamic spectra side by side. Most striking is the faint pulses occurring in both epochs: 18-1A, 18-2A, and 19-1A, B, C, 19-2A, B, C -- these pulses all have very similar drift rates ($\sim1.3-1.4\,\mathrm{MHz}\,\mathrm{s}^{-1}$, virtually indistinguishable within the bandwidth of our observations), durations ($\sim 12$\,min), and are similar in their average flux densities ($\sim 40-50$\,mJy). However, in the 2019-03-07 epoch, three faint pulses with equal drift rates occur every rotation, where as in 2018-10-02 only one faint, negative-drifting pulse occurs per rotation. Pulses 19-1B and 19-2B are morphologically most similar to the faint pulses 18-1A and 18-2A occurring once per rotation in the 2018-10-02 epoch, 18-1A and 18-2A. Furthermore, the time difference between 19-1B and 19-1D, and 19-2B and 19-2D ($\sim 60$\,min) is similar to the time difference between 18-1A and 18-1B ($\sim 50$\,min).

While there are some morphological similarities between the bright pulses in both epochs, the bright burst in 2018-10-02 has a slower bulk drift rate, a longer duration, and overall a more complex morphology to the bright pulses present in 2019-03-07. Nonetheless, similarities between the pulses in both epochs suggests that some of the pulses originate from source regions that have remained relatively stable over timescales of several months. This suggestion was also made by \citet{Villadsen} to explain the similar morphologies of several bursts detected from UV~Ceti, despite months-long separation between observations. In particular, we suggest that the similar morphology, flux density, and drift rate of pulses 18-1/18-2A and 19-1/19-2B, along with their similar time gap from the subsequent bright pulses 18-1B, and 19-1/19-2D, indicates that they originate from the same source region along a stable magnetic loop. 

Under the assumption that they are the from same source region, and that their host magnetic loop has not drifted significantly in magnetic longitude, we can set the peak times of 18-1A and 19-1B as our reference time for zero longitude for the 2018-10-02 and 2019-03-07 epochs, respectively. We use the $5.447\pm 0.008\,$h periodicity measured from the 2019-03-07 epoch to calculate the relative magnetic longitude from the set reference time. This enables us to more closely analyse the similarities and differences of the sets of pulses we detect in our two observing epochs. Relative magnetic longitudes are shown in the upper abscissae of Figures \ref{fig:uvceti_lc}, \ref{fig:ds_2018}, \ref{fig:2018_DS_IQ}, \ref{fig:ds_2019}, \ref{fig:ds_2019_substructure}, and \ref{fig:compare_bursts}. 

Figure \ref{fig:compare_bursts} gives a comparison of the pulses 18-1A, B with 19-1A, B, C and 19-2A, B, C. Dashed lines in this figure indicate the location and drift rates of the pulses. This figure highlights the similar drift rates of the faint pulses in both epochs. The evolution in drift rate and relative magnetic longitude from 18-1B to 19-1D and 19-1D is also evident, along with the quasi-periodic nature of pulses 19-1A, B, C and 19-2A, B, C.


\section{DISCUSSION}


    \subsection{\label{sec:emission_origin}Origin of the emission}
    Incoherent emission has a maximum brightness temperature of $10^{11}-10^{12}$\,K due to inverse Compton scattering \citep{Kellerman}, and so any radio emission with a brightness temperature persistently exceeding this limit must be coherent in nature (disregarding relativistic Doppler boosting). We used Equation 14 from \citet{Dulk} to calculate the brightness temperature $T_b$ as follows:
    \begin{equation}
        T_b = \frac{1}{2k_B}S_\nu \left(\frac{c}{\nu} \right)^2\left(\frac{d}{l}\right)^2,
    \end{equation}
    where $k_B$ is the Boltzmann constant, $S_\nu$ is the flux density in Jy, $c$ is the speed of light, $\nu$ is the observing frequency, $d$ is the distance to the source, and $l$ is the emission region size. Assuming the radio emission is beamed and rotationally modulated (see Section \ref{sec:burst_period}) we can use the projected rotational velocity of $32.2\,\mathrm{km}\,\mathrm{s}^{-1}$ of UV~Ceti \citep{UVCetRotation}, and an upper limit on the duration of all pulses of $20$\,min, to set an upper limit on the emission region size for each pulse of $3.6\times10^{9}\,\mathrm{cm} = 0.50\,R_{\mathrm{Jup}}$. Using this upper limit on the source size, we calculated a lower limit of the brightness temperature on the pulses from both epochs.
    
    We find that the brightness temperature lower limits are in the range of $0.22-4.3\times10^{13}\,\mathrm{K}$, exceeding the $10^{12}\,\mathrm{K}$ inverse-Compton limit \citep{Kellerman}, therefore indicating that the emission is coherent. This conclusion is also supported by the high degree of circular and elliptical polarization (up to 100 per~cent), and the complex temporal and structure of the pulses.The periodicity shows that the pulses originate from persistent, highly beamed sources whose emission crosses our line of sight as the star rotates, confirming suggestions by \citet{Lynch} and \citet{Villadsen}.
    
    The two emission mechanisms that are generally invoked to explain stellar coherent radio bursts are plasma emission \citep{Dulk} and the electron cyclotron maser instability (ECMI; \citealp{Treumann}).
    The ECMI operates most efficiently when $\nu_p/\nu_c \ll 1$ i.e. in low-density, highly magnetized plasmas. Here, $\nu_c$ is the electron cyclotron frequency, given by
    \begin{equation}
    \label{eq:cyclotron_freq}
        \nu_c = eB/2\pi m_e c \approx 2.8 B\,\mathrm{MHz}.
    \end{equation}
    The emission is expected to be very narrow-band, centred around the electron cyclotron frequency
    and possibly its harmonics. In favourable conditions for the ECMI, when the ratio of the plasma frequency
    \begin{equation}
        \nu_p = \sqrt{n_e e^2/\pi m_e} \approx 9 \sqrt{n_e}\,\mathrm{kHz},
    \end{equation}
    to the cyclotron frequency is small (i.e. $\nu_p/\nu_c \ll 1$), the emission is expected to be polarized in the free-propagating extraordinary mode of the birefringent plasma ($x$-mode), and is beamed at angles nearly perpendicular to the magnetic field \citep{Treumann}. However, the ECMI can also operate with reduced efficiency when a weaker condition, $\nu_p/\nu_c \leq 1$, is met. In this instance, polarization in the ordinary mode ($o$-mode) is possible. When the line-of-sight component of the magnetic field is directed towards the observer, then $x$-mode corresponds with RCP and $o$-mode with LCP. In the opposite orientation, $x$-mode corresponds with LCP and $o$-mode with RCP. Therefore, with knowledge of the magnetic field geometry, the observed sense of polarization can be used to determine the magneto-ionic mode of the emission.

    Plasma emission occurs at the local plasma frequency and its second harmonic; higher frequencies/harmonics are expected to be absorbed via gyroresonance or free-free absorption \citep{Dulk}. Plasma emission at the fundamental mode is expected to be polarized in the $o$-mode, and  mildly polarized at the second harmonic \citep{Melrose78, Melrose93}.

    In Section \ref{sec:linearpol}, we use the presence of linear polarization in the pulses to place an upper limit on the electron number density of $n_e \lesssim 41\,\mathrm{cm}^{-3}$.
    This density limit corresponds to plasma frequencies less than 54\,kHz, well below our observing frequency. Furthermore, all the detected pulses in 2019-03-07 are RCP, indicating emission consistently from either the $x$ or $o$-mode. \citet{UVCetiZDI} showed that the magnetic north pole of UV~Ceti is inclined towards Earth, meaning that the emission is most likely to be from the northern hemisphere of the UV~Ceti magnetosphere i.e. where the line-of-sight magnetic field component is directed towards the observer. This indicates that the emission is in the $x$-mode.
    
    Elliptical polarization is also indicative of radiation in the $x$-mode, as argued in the case of elliptically polarized Jovian decametric radiation \citep{Melrose91, Melrose93, Dulk94}. The presence of elliptically polarized emission in both epochs gives further evidence that the emission is emitted in the $x$-mode. 
    
    Together, polarization properties of the pulses in both epochs favour emission in the $x$-mode from the northern magnetic hemisphere of UV~Ceti, showing that the emission is from the ECMI. This is consistent with recent detections of bursts from UV~Ceti by \citet{Lynch} and \citet{Villadsen}, and more broadly with the nature of periodic radio pulses from other auroral emitters (e.g. \citealp{Lynch15, Hallinan07, CUVir, Zarka98}). 
    
    Using Equation \ref{eq:cyclotron_freq}, and assuming emission in the second harmonic, the local magnetic field strength at the emission regions is in the range of $133-184$\,G (or $266-368$\,G if the emission is in the fundamental mode). Assuming a purely dipolar magnetospheric geometry with mean surface magnetic field strength of 6.7\,kG as a first order approximation, based on Zeeman Doppler Imaging of UV~Ceti, \citep{UVCetiZDI}, then this implies source altitudes of $\sim 3.3-3.7\,R_*$ (or $2.6-2.9\,R_*$ if the emission is in the fundamental mode), where $R_*$ is the stellar radius ($1R_* = 0.159\pm 0.006\,R_\odot$; \citealp{kervella16}). 

    
    
    \subsection{Origin of frequency drift}
    Wide-band dynamic spectra of the ultra-cool dwarfs (UCDs) TVLM\,513$-$46, and 2M\,0746$+$20 presented by \citet{Lynch15} exhibited periodic radio pulses with a linear frequency drift, similar to our observations of pulses from UV~Ceti. \citet{Lynch15} explained that the drifting features in their dynamic spectra are caused by ECMI emission along a few localised magnetic loops in the stellar magnetospheres. This frequency drift arises due to the fact that ECMI emission emits in a narrow frequency range about the electron cyclotron frequency, determined by the local magnetic field strength (see Equation \ref{eq:cyclotron_freq}). Source regions at different altitudes along an active magnetic loop will have differing local magnetic field strengths and therefore different emission frequencies. As the star rotates, the frequency-dependent ECMI emission beam will cross the line of sight at different orientations which depend on the emission altitude, and therefore on observing frequency. 
    We suggest that a similar geometrical effect is responsible for the observed frequency drifts in our dynamic spectra, although a detailed modelling procedure of the magnetospheric geometry responsible for the pulses is beyond the scope of this work.
    
    

    \subsection{\label{sec:linearpol}Interpretation of linear polarization}
  
    \citet{Melrose91} discussed possible causes for the linear polarization in Jovian decametric emission observed e.g. by \citet{JupiterDAM} and \citet{Boudjada}. They concluded that the presence of elliptically polarized emission implies extremely low plasma densities in the emission region. Equation 14 from \citet{Melrose91} gives an upper limit on the electron density when the emission is elliptically polarized:
    \begin{equation}
        n_e \lesssim \alpha (\nu/ 25\,\mathrm{MHz}),
    \end{equation}
    where $\alpha$ is a geometrical factor of order unity and $\nu$ is the observing frequency. For our highest observing frequency (1030\,MHz), and taking $\alpha = 1$, this gives an approximate upper limit on the electron density of $41\,\mathrm{cm}^{-3}$. 
    
    We can compare this density to an order-of-magnitude estimate of the expected density at a source altitude of $\sim 3.7\,R_*$ (see Section \ref{sec:emission_origin}) as follows. We assume a simple exponential coronal density model, adopting a scale height of $0.48\,R_*$ calculated by \citet{Villadsen} using a constant-gravity hydrostatic equilibrium model, and take an average coronal base density of $10^{10.5}\,\mathrm{cm}^{-3}$ (similar to densities inferred for other active M-dwarfs; \citealp{Ness}). This gives an expected electron density on the order of $10^8\,\mathrm{cm}^{-3}$ at the source altitudes.
    
    Our upper limit of $n_e \lesssim 41\,\mathrm{cm}^{-3}$ is much lower than this order-of-magnitude estimate of the electron density, and is unlikely to be explained by permitting a more complex model (e.g. removing the assumption that the magnetosphere is strictly dipolar). A plausible explanation for our strong upper limit on electron density is that the emission arises from an extreme density cavity, analogous to the cavities present around the emission regions of Earth's auroral kilometric radiation (AKR), generated by powerful magnetic field-aligned electric fields \citep{Zarka98, Ergun}. \citet{Melrose91} concluded that densities low enough to permit elliptically polarized emission from active stars or the Sun are highly unlikely to occur due to high average coronal electron densities, highlighting the remarkable nature of this emission.
    
    Although ECMI has often been the favoured mechanism to explain stellar coherent bursts, a common difficulty encountered is the expected absorption of ECMI emission by the overlying plasma in the gyroresonance harmonic layers \citep{Dulk}. Assuming a dipolar magnetosphere and emission in the second harmonic, our emission region is at altitudes between $\sim3.3-3.7\,R_*$ (see Section \ref{sec:emission_origin}) -- a radial extent of $\sim 0.4\,R_*$. The third gyroresonance layer, where $B =  2 B_\mathrm{src}/3$, spans altitudes of $\sim3.8-4.2\,R_*$. The presence of elliptical polarisation across the entire bandwidth of some pulses requires that the associated cavity extends at least $0.4\,R_*$. It therefore seems plausible that the cavity may extend an additional $0.5\,R_*$ in altitude, such that it reaches the third gyroresonance harmonic layer for the lowest emission frequency. This is supported by the observation of elliptically polarized emission from UV~Ceti by \citet{Lynch} at 154\,MHz, which corresponds to a dipolar magnetospheric altitude of $\sim 6.2\,R_*$. If the inferred cavity does extend to the third, or possibly even the second gyroresonance harmonic layer (where $B = B_\mathrm{src}/2$, which strongly absorbs emission in the fundamental mode), then the extremely low densities should result in an optical depth sufficiently small for the emission to escape.

    Earth's auroral kilometric radiation (AKR) has been observed to escape the plasma via multiple reflections from within the auroral density cavities, to a height where the ambient $x$-mode cutoff frequency is equal to the emission frequency (e.g. \citealp{Zarka98, Ergun}). The extremely low-density cavity inferred by the presence of linear polarization should be surrounded by the high-density plasma of the stellar corona, suggesting that ECMI emission observed here may also escape the stellar magnetosphere through a similar process. These reflections would be likely to modify the intrinsically elliptically polarized emission via mode-coupling, potentially removing any linearly polarized component of the emission. A possible solution is to require that the cavity has a sufficiently large volume so that radiation can escape without reflecting off the cavity walls. Another solution may be that the elliptical polarization is generated by partial conversion of $x$-mode into $o$-mode after reflections on the cavity wall \citep{Zarka98}, removing any strong requirements on the minimum size of the cavity. We note that elliptically polarized emission has also been observed from AD~Leo \citep{Spangler74}, suggesting that the auroral processes generating these magnetospheric cavities may also occur in other active M-dwarfs. 
    
    Finally, assuming intrinsically elliptically polarized emission, we can use the observed elliptical polarization to determine the emission angle of the ECMI emission relative to the local magnetic field. \citet{Melrose93} gave a simple relation between the ECMI emission angle relative to the magnetic field, $\theta$, and the axial ratio of elliptically polarized emission, $T$ (not to be confused with temperature, or brightness temperature $T_b$), to first order:
    \begin{equation}
        T \approx \cos(\theta).
    \end{equation}
    Using the expression for the axial ratio from \citet{Dulk94},
    \begin{equation}
        T = \tan\left(\frac{1}{2}\arcsin\left(\frac{V}{I_p}\right)\right),
    \end{equation}
    where $I_p = \sqrt{Q^2 + U^2 + V^2}$, we can derive the angle of ECMI emission for our 2019-03-07 observation, where full polarization calibration was possible. For pulses 19-1B and 19-2C, which are detected significantly in linear polarization, we derive an ECMI emission angles of $60\pm4^\circ$ and $58\pm5^\circ$ to the magnetic field respectively. 
    
    

    \subsection{\label{sec:auroral}Transition from solar-type to auroral activity}
    The multi-wavelength flare activity of UV~Ceti (along with other active M-dwarfs) has been largely interpreted within the framework of solar flares. For example, the observations of UV~Ceti performed by \citet{Guedel96}, described in Section~\ref{sec:intro}, confirmed the operation of the Neupert effect on active M-dwarfs. The Neupert effect is believed to be driven by the process of chromospheric evaporation, where accelerated non-thermal electrons stream downwards along magnetic loops, impacting and heating the chromospheric plasma, which then `evaporates' and rises to the upper corona. 
    Chromospheric evaporation has been observed in other active M-dwarfs (e.g. \citealp{proxcen_evap}), suggesting that Solar-like magnetospheric processes may be prevalent among active M-dwarfs, although several counter-examples exist (e.g. \citealp{Osten05, Lim96}).
    
    
    Another paradigm of magnetospheric activity that may be relevant to active M-dwarfs is auroral-type magnetic activity. Auroral processes, revealed by periodic, highly polarized radio pulses, are known to occur on magnetized planets \citep{Zarka98}, brown dwarfs \citep{Hallinan07, Pineda}, and magnetic chemically peculiar stars \citep{CUVir}. Auroral processes are driven by large, powerful  field-aligned current systems occurring in stellar and sub-stellar magnetospheres \citep{Treumann, Zarka98}. Our detection of periodic, elliptically polarized pulses from UV~Ceti, confirming suggestions by \citet{Lynch} and \citet{Villadsen}, shows that this type of magnetic activity also applies to active M-dwarfs. This shows that auroral emitters extend further up along the main sequence than previously realised. 
    
    Our detection of elliptically polarized emission from UV~Ceti (and previously by \citealt{Lynch}) implies that extreme density cavities exist within the magnetosphere of UV~Ceti (see Section \ref{sec:linearpol}). This shows that the auroral processes operating in the magnetosphere of UV~Ceti are closely analogous those occurring in the magnetosphere of Jupiter and Saturn, where elliptically polarized emission from extremely low densities is also observed \citep{ Melrose91, JupiterDAM, Boudjada, fischer}. Elliptically polarized emission was also detected from AD~Leo by \citet{Spangler74}, which, like UV~Ceti, hosts a large-scale, axisymmetric magnetic field \citep{UVCetiZDI, Morin08}, hinting that auroral-type magnetic activity may apply to other active M-dwarfs with large, axisymmetric magnetic fields. We note, however, that no other M-dwarf has shown evidence of periodic radio emission \citep{Villadsen}. This indicates that such behaviour may only be intermittent, or it could also be due to incomplete observational coverage of the rotational periods of these stars (e.g. AD~Leo has rotational period of $2.2399\pm0.0006$\,d; \citealp{Morin08}).
    
    Together, the observation of both Solar-like \citep{Guedel96} and auroral activity in UV~Ceti, and potentially other active M-dwarfs \citep{Spangler74}, suggests that these active stars may mark the beginning of the transition from Solar-like activity to auroral activity that becomes prevalent in substellar objects \citep{Pineda, Zarka98}. Further detailed studies of coherent bursts from active M-dwarfs across a range of spectral types and magnetospheric geometries, along with simultaneous multi-wavelength information may assist in clarifying the stellar parameters and magnetospheric conditions which drive this transition.
    

\section{Summary \& Conclusions}
We report the detection of several coherent, highly polarized pulses from the active M-dwarf UV~Ceti. The pulses are periodic in nature, with a period of $5.447\pm 0.008$\,h -- consistent with the measured rotational period of UV~Ceti \citep{UVCetRotation}. The brightness temperature, polarization, and time-frequency structure of the pulses are consistent with highly-beamed ECMI emission originating along magnetic loops, which cross the line of sight every rotation of the star. Some of the pulses show elliptical polarization -- a rare occurrence in nature, which implies very low plasma densities ($\lesssim 41\,\mathrm{cm}^{-3}$) around the source region. This can be explained if the emission originates from an extreme density cavity in the magnetosphere of UV~Ceti, analogous to the terrestrial auroral cavities \citep{Zarka98, Ergun}. We suggest that the density cavity is key to the escape of the radiation at the first and second harmonic of the cyclotron frequency, by a reduced optical depth at the gyroresonance harmonic layers, and/or by multiple reflections along the cavity wall. The presence of periodic and elliptically polarized radio emission from UV~Ceti shows that auroral-type magnetic activity can also manifest in active M-dwarfs, along with Solar-like activity as indicated by the presence of the Neupert effect \citep{Guedel96}. This indicates that M-dwarfs may mark the beginning of the transition from Solar-like activity to auroral activity which becomes prevalent in brown dwarfs and magnetized planets \citep{Pineda, Zarka98}.

We have demonstrated the capabilities of ASKAP for studying polarized variable radio sources.  
    We note that our best RMS noise in our 4\,MHz, 10\,s dynamic spectrum bins is $\sim 7$\,mJy. This RMS noise value is competitive with the RMS noise for dynamic spectra taken with the Karl G. Jansky Very Large Array (1-7\,mJy) reported by \citet{Villadsen}, with similar spectral and temporal resolution. 
    
    The quality of the dynamic spectra, and the wide field of view of ASKAP presents a strong opportunity for future large-scale surveys for coherent bursts from stellar and sub-stellar objects, along with detailed, targeted studies as presented in this work. \citet{Villadsen} estimated that the active M-dwarf transient density should be one per 10 deg$^2$, meaning ASKAP surveys such as VAST \citep{Murphy13} will probe the population of coherent bursts from M-dwarfs, giving a more complete insight into the processes operating in the magnetospheres of these stars. 

\section*{Acknowledgements}
We thank Donald Melrose, Mohammad Rafat, Michael Wheatland and George Heald for useful discussions and suggestions which strengthened this work.
TM acknowledges the support of the Australian Research Council through grant FT150100099. AZ acknowledges support from an Australian Government Research Training Program (RTP) Scholarship. DK was supported by NSF grant AST-1816492.
The Australian SKA Pathfinder is part of the Australia Telescope National Facility which is managed by CSIRO. Operation of ASKAP is funded by the Australian Government with support from the National Collaborative Research Infrastructure Strategy. ASKAP uses the resources of the Pawsey Supercomputing Centre. Establishment of ASKAP, the Murchison Radio-astronomy
Observatory and the Pawsey Supercomputing Centre are initiatives of the Australian Government, with support from the Government of Western Australia and the Science and Industry Endowment Fund. We acknowledge the Wajarri Yamatji people as the traditional owners of the Observatory site. Parts of this research were supported by the Australian Research Council Centre of Excellence for All Sky Astrophysics in 3 Dimensions (ASTRO 3D), through project number CE170100013.
This research made use of the following \textsc{python} software packages: \textsc{aplpy}\footnote{\url{http://aplpy.github.io/}}, an open-source plotting package for \textsc{python} \citep{aplpy}; \textsc{matplotlib}\footnote{\url{https://matplotlib.org/}} \citep{matplotlib}; \textsc{numpy}\footnote{\url{https://www.numpy.org/}} \citep{numpy}; and \textsc{astropy}\footnote{\url{http://www.astropy.org}}, a community-developed core \textsc{python} package for Astronomy \citep{astropy:2013, astropy:2018}.

\bibliographystyle{mnras}
\bibliography{askapuvceti}

\end{document}